\documentclass[a4paper,twocolumn,11pt,unpublished]{quantumarticle}
\pdfoutput=1
\usepackage[utf8]{inputenc}
\usepackage[english]{babel}
\usepackage[T1]{fontenc}
\usepackage{amsmath}
\usepackage{hyperref}

\usepackage{tikz}
\usepackage{lipsum}

\begin{document}

\title{A no-go result on observing quantum superpositions}

\author{Guang Ping He}
\affiliation{School of Physics, Sun Yat-sen University, Guangzhou 510275, China}
\email{hegp@mail.sysu.edu.cn}
\orcid{0000-0002-3541-409X}
\maketitle

\begin{abstract}
We give a general proof showing that once irreversible processes are
involved, a class of projective measurements is impossible. Applying this
no-go result to the Schr\"{o}dinger's cat paradox implies that if something
is claimed to be a real Schr\"{o}dinger's cat, there will be no measurable
difference between it and a trivial classical mixture of ordinary cats in
any physically implementable process, otherwise raising the dead will become
reality. Other similar macroscopic quantum superpositions cannot be observed either due to the lack of non-commuting measurement bases. Our proof does not involve any quantum interpretation theory and
hypothesis.
\end{abstract}


\section{Introduction}
Schr\"{o}dinger's cat \cite{qi2222,qi2223}, i.e., the outcome of a
gedankenexperiment which is supposed to be a quantum superposition of both a
living cat and a dead one, is undoubtedly one of the most fascinating
paradoxes of quantum mechanics. Over the years, there were many experimental
implementations of the so-called cat states using lifeless objects \cite%
{qi2257,qi2258,qi2259,qi2261,qi2235,qi2248,qi2254,qi2256,qi2243,qi2244,qi2242,qi2228,qi2229}%
, with more possible implementations \cite%
{qi2221,qi2220,qi2251,qi2245,qi2219,qi2250,qi2249,qi2252,qi2265} on the way.
But we still have not found such a superposition on a real cat or living
objects. Is it possible that someday we may eventually be able to find a
real Schr\"{o}dinger's cat as technology advances? Here it will be shown
that any such attempt will be futile. To be specific, we will prove that a
class of projective measurements is impossible once irreversible processes
are involved. Applying this no-go proof to the Schr\"{o}dinger's cat paradox
implies that even if something is claimed to be a real Schr\"{o}dinger's
cat, it can never be verified physically. There will be no measurable
difference between this state and a trivial classical mixture of living cats
and dead cats in any physical process. This result answers why quantum
superpositions of living objects have never been found, and predicts that
they never will. Our no-go proof bases strictly on currently well-accepted
physics laws without involving any quantum interpretation theory and
hypothesis. It can also be generalized to cover other quantum states related
with irreversible processes.

\section{A general no-go result}

Our key idea is simple. Keep in mind that quantum mechanics is not the only
science that governs the universe. At least, there is also thermodynamics,
whose second law tells us that many processes in our world are irreversible.
Note that in thermodynamics, an irreversible process usually implies that
the process cannot be reversed along the same path, as widely found in
isolated systems. But here we focus our discussion on a special set of these
processes, in which the final state cannot be returned to the initial state
along any path in a given laboratory (the meaning of this \textquotedblleft
laboratory\textquotedblright\ will be further explained after the proof of
the theorem). When such a process is involved, we have the following general
result.

\bigskip

\textbf{The no-observing theorem:}

\textit{If in a given laboratory: (i) the states $\left\vert L\right\rangle $ and $%
\left\vert D\right\rangle $ can be discriminated unambiguously, and (ii) the
evolution from $\left\vert D\right\rangle $ to $\left\vert L\right\rangle $\
is impossible (i.e., there does not exist any operation $T$ such that $%
T\left\vert D\right\rangle \rightarrow \left\vert L\right\rangle $, even
with a successful probability less than $1$), then in this laboratory the
following projective operator is impossible:%
\begin{equation}
P_{S}=(a\left\vert L\right\rangle +b\left\vert D\right\rangle )(a^{\ast
}\left\langle L\right\vert +b^{\ast }\left\langle D\right\vert )  \label{Ps}
\end{equation}%
for any $a$ and $b$ that satisfy $\left\vert a\right\vert ^{2}+\left\vert
b\right\vert ^{2}=1$, $a\neq 0$\ and $b\neq 0$.
}

\bigskip

\textbf{Proof:}

The condition (i) actually implies that $\left\vert L\right\rangle $ and $%
\left\vert D\right\rangle $ are orthogonal to each other, and there exists a
projective operator%
\begin{equation}
P_{L}=\left\vert L\right\rangle \left\langle L\right\vert
\end{equation}%
which can tell the two states apart. Now assume that the operator $P_{S}$\
in Eq. (\ref{Ps}) exists. Then for a system initialized on the state $%
\left\vert D\right\rangle $, applying $P_{S}$ on it stands a probability $%
\left\vert b\right\vert ^{2}$ to project successfully because $%
P_{S}\left\vert D\right\rangle =b^{\ast }(a\left\vert L\right\rangle
+b\left\vert D\right\rangle )$. When the projection is successful, the final
state is%
\begin{equation}
\left\vert S\right\rangle =a\left\vert L\right\rangle +b\left\vert
D\right\rangle .
\end{equation}%
Further applying the projective operator $P_{L}$ on it also stands a
probability $\left\vert a\right\vert ^{2}$ to be successful because $%
P_{L}\left\vert S\right\rangle =a\left\vert L\right\rangle $, and the final
state becomes $\left\vert L\right\rangle $. Thus, through the combination of
the operations $P_{S}$\ and $P_{L}$, the state $\left\vert D\right\rangle $
is eventually turned into $\left\vert L\right\rangle $ with probability $%
\left\vert a\right\vert ^{2}\left\vert b\right\vert ^{2}$. This result
conflicts with the premise that the evolution from $\left\vert
D\right\rangle $ to $\left\vert L\right\rangle $\ is impossible in this
laboratory, unless either $a$ or $b$ equals $0$. Therefore, the projective
operator $P_{S}$ cannot exist in this laboratory for any nonvanishing $a$
and $b$. 

\bigskip

\textbf{Remarks:}

(1) In fact, even if the projection on the state $\left\vert D\right\rangle $
using $P_{S}$ fails, it is equivalent to applying the operator $I-P_{S}$ and
the projection is successful. In this case, the resultant state can be
written as%
\begin{equation}
\left\vert S^{\prime }\right\rangle =a^{\prime }\left\vert L\right\rangle
+b^{\prime }\left\vert D\right\rangle
\end{equation}%
where $a^{\prime }$\ and $b^{\prime }$ ensure that $\left\vert S^{\prime
}\right\rangle $\ is orthogonal to $\left\vert S\right\rangle $. Further
applying $P_{L}$ on $\left\vert S^{\prime }\right\rangle $\ stands a
probability $\left\vert a^{\prime }\right\vert ^{2}$ to be successful, so
that the final state is $\left\vert L\right\rangle $ too. Thus, $\left\vert
a\right\vert ^{2}\left\vert b\right\vert ^{2}$ is only a loose lower bound
of the total probability for turning $\left\vert D\right\rangle $ into $%
\left\vert L\right\rangle $, which is sufficient for our proof. Moreover, if
we are still not satisfied with the successful probability, we can even
repeat the measurements $P_{S}$\ and $P_{L}$ many times, because once the
two measurements fail to turn the state into $\left\vert L\right\rangle $,
the final state will always return to the initial state $\left\vert
D\right\rangle $, so that the above procedure can be applied again, and the
probability for turning $\left\vert D\right\rangle $ into $\left\vert
L\right\rangle $ can be made arbitrarily close to $1$.

(2) The \textquotedblleft laboratory\textquotedblright\ in the theorem means
the scale on which the states $\left\vert L\right\rangle $ and $\left\vert
D\right\rangle $ satisfy the two conditions in the theorem. It can be either
an isolated system or an open one. It can also be a real laboratory, or even
the universe. If for the two distinguishable states $\left\vert
L\right\rangle $ and $\left\vert D\right\rangle $, the operation for turning
$\left\vert D\right\rangle $ to $\left\vert L\right\rangle $\ is beyond the
technology of a specific real laboratory, then the theorem implies that the
operator $P_{S}$\ is impossible in this laboratory. Or if the evolution from
$\left\vert D\right\rangle $ to $\left\vert L\right\rangle $\ is impossible
in the whole universe, then the operator $P_{S}$\ cannot exist in our world
at all.

(3) An irreversible process from $\left\vert L\right\rangle $ to $\left\vert
D\right\rangle $ generally means that the evolution $\left\vert
L\right\rangle \rightarrow \left\vert D\right\rangle $\ is possible, while
only\ $\left\vert D\right\rangle \rightarrow \left\vert L\right\rangle $ is
forbidden. But from the proof of the theorem we can see that it does not
require the existence of $\left\vert L\right\rangle \rightarrow \left\vert
D\right\rangle $. Therefore, the theorem not only applies to irreversible
processes, but also applies to any two completely irrelevant states $%
\left\vert L\right\rangle $ and $\left\vert D\right\rangle $, i.e., both the
evolutions $\left\vert L\right\rangle \rightarrow \left\vert D\right\rangle $%
\ and\ $\left\vert D\right\rangle \rightarrow \left\vert L\right\rangle $
can be absent. For example, if turning stones to bread is impossible in a
given laboratory, nor vice versa, then the operator $(a\left\vert
stone\right\rangle +b\left\vert bread\right\rangle )(a^{\ast }\left\langle
stone\right\vert +b^{\ast }\left\langle bread\right\vert )$\ is impossible
for any nonvanishing $a$ and $b$ in this laboratory either. In other words,
the theorem can be evaded only if the evolution between the two states $%
\left\vert L\right\rangle $ and $\left\vert D\right\rangle $ is reversible,
i.e., both the operations $T$ and $T^{\prime }$\ must exist such that\ $%
T\left\vert D\right\rangle \rightarrow \left\vert L\right\rangle $ and $%
T^{\prime }\left\vert L\right\rangle \rightarrow \left\vert D\right\rangle $%
, at least with a non-vanishing probability.

\section{Schr\"{o}dinger's cats vs classical mixtures}

Now let us consider the Schr\"{o}dinger's cat. According to the author's
original description \cite{qi2222}, a cat is penned up in a chamber, along
with a device in which a bit of radioactive substance decays with
probability $50\%$ after an hour. If it decays, a hammer will be triggered
which shatters a small flask of hydrocyanic acid and poisons the cat to
death. Otherwise, the cat remains alive. Let $\left\vert
decayed\right\rangle $ ($\left\vert undecayed\right\rangle $) denote the
final state of the device when the radioactive substance decayed (did not
decay). The above description suggests that the wavefunction of the device
and the cat at the end of the process is%
\begin{equation}
\left\vert \Psi _{S}^{+}\right\rangle =\frac{1}{\sqrt{2}}\left( \left\vert
undecayed\right\rangle \left\vert alive\right\rangle +\left\vert
decayed\right\rangle \left\vert dead\right\rangle \right)  \label{sch}
\end{equation}

For comparison, we also consider a classical mixture with trivial
significance. Suppose that $50$ living cats and $50$ dead ones were put in
an opaque room, and we randomly grab one out. The probabilities for grabbing
a living one or a dead one are both $50\%$. This ensemble is not a real Schr%
\"{o}dinger's cat because each single cat in the room is in a deterministic
classical state. In the language of quantum mechanics, it is a classical
mixture described by the density matrix%
\begin{equation}
\rho _{cat}=\frac{1}{2}\left\vert alive\right\rangle \left\langle
alive\right\vert +\frac{1}{2}\left\vert dead\right\rangle \left\langle
dead\right\vert .  \label{mixture}
\end{equation}%
There is no coherence between the states $\left\vert alive\right\rangle $
and $\left\vert dead\right\rangle $, i.e., the phase difference between them
does not take any fixed value. Consequently, it is not a quantum
superposition of the two states on a single individual.

In Eq. (\ref{sch}), if we measure the state of the cat only (suppose that we
have many copies of the same state and perform the measurement many times),
it will show no difference from the above classical mixture, because the
reduce density matrix of the cat in Eq. (\ref{sch}) is $tr\left\vert \Psi
_{S}^{+}\right\rangle \left\langle \Psi _{S}^{+}\right\vert $ (taking
partial trace over all possible states of the device), which equals exactly
to Eq. (\ref{mixture}).

On the other hand, suppose that we measure the state of the cat and the
device collectively in the basis $\{\left\vert undecayed\right\rangle
\left\vert alive\right\rangle ,\left\vert decayed\right\rangle \left\vert
dead\right\rangle \}$. (In fact, a complete basis for this composite system
should also include two more states, e.g., $\left\vert
undecayed\right\rangle \left\vert dead\right\rangle $ and $\left\vert
decayed\right\rangle \left\vert alive\right\rangle $, or the superpositions
of these two states. But they are not important to our discussion, because
they will never be found as the result of the measurement. Therefore, for
simplicity, in the following we will not mention these two additional states
when speaking of this basis.) The outcome will be either $\left\vert
undecayed\right\rangle \left\vert alive\right\rangle $ or $\left\vert
decayed\right\rangle \left\vert dead\right\rangle $\ with equal
probabilities $50\%$. This result does not necessarily lead to any paradox
either, because there also exists another classical mixture which can
display exactly the same behavior. That is, consider that there is an
ensemble of $100$ chambers. With the equal probabilities, each chamber
contains either a living cat and a device with undecayed radioactive
substance, or a dead cat and a device with decayed radioactive substance.
Then it is a classical mixture described by the density matrix%
\begin{eqnarray}
\rho _{S} &=&\frac{1}{2}\left\vert undecayed\right\rangle \left\vert
alive\right\rangle \left\langle alive\right\vert \left\langle
undecayed\right\vert  \nonumber \\
&&+\frac{1}{2}\left\vert decayed\right\rangle \left\vert dead\right\rangle
\left\langle dead\right\vert \left\langle decayed\right\vert .
\label{classical}
\end{eqnarray}%
Measuring it in the basis $\{\left\vert undecayed\right\rangle \left\vert
alive\right\rangle ,\left\vert decayed\right\rangle \left\vert
dead\right\rangle \}$ will also result in $\left\vert undecayed\right\rangle
\left\vert alive\right\rangle $ or $\left\vert decayed\right\rangle
\left\vert dead\right\rangle $\ with the same probabilities.

What really makes the Schr\"{o}dinger's cat interesting is that if we have
the projective measurement operator%
\begin{equation}
P_{Sch}^{+}=\left\vert \Psi _{S}^{+}\right\rangle \left\langle \Psi
_{S}^{+}\right\vert  \label{project both}
\end{equation}%
and apply it on the entire composite system containing both the device and
the cat, then the measurement result will always be $\left\vert \Psi
_{S}^{+}\right\rangle $ with probability $100\%$ because $%
P_{Sch}^{+}\left\vert \Psi _{S}^{+}\right\rangle =\left\vert \Psi
_{S}^{+}\right\rangle $. On the contrary, Eq. (\ref{classical}) can also be
expressed as%
\begin{equation}
\rho _{S}=\frac{1}{2}\left\vert \Psi _{S}^{+}\right\rangle \left\langle \Psi
_{S}^{+}\right\vert +\frac{1}{2}\left\vert \Psi _{S}^{-}\right\rangle
\left\langle \Psi _{S}^{-}\right\vert
\end{equation}%
where we define%
\begin{equation}
\left\vert \Psi _{S}^{-}\right\rangle =\frac{1}{\sqrt{2}}\left( \left\vert
undecayed\right\rangle \left\vert alive\right\rangle -\left\vert
decayed\right\rangle \left\vert dead\right\rangle \right) .
\end{equation}%
Applying $P_{Sch}^{+}$ on $\rho _{S}$ will project it to either $\left\vert
\Psi _{S}^{+}\right\rangle $ or $\left\vert \Psi _{S}^{-}\right\rangle $\
with the equal probabilities $50\%$. That is, the projective measurement $%
P_{Sch}^{+}$ can discriminate the quantum superposition Eq. (\ref{sch}) from
the classical mixture Eq. (\ref{classical}).

Furthermore, other than using $P_{Sch}^{+}$ on Eq. (\ref{sch}), we can
measure the device in the basis $\{\left\vert \phi _{dev}^{+}\right\rangle
,\left\vert \phi _{dev}^{-}\right\rangle \}$ first, where%
\begin{equation}
\left\vert \phi _{dev}^{\pm }\right\rangle =\frac{1}{\sqrt{2}}\left(
\left\vert undecayed\right\rangle \pm \left\vert decayed\right\rangle
\right) .
\end{equation}%
Since Eq. (\ref{sch}) can also be expressed as%
\begin{equation}
\left\vert \Psi _{S}^{+}\right\rangle =\frac{1}{\sqrt{2}}\left( \left\vert
\phi _{dev}^{+}\right\rangle \left\vert \psi _{cat}^{+}\right\rangle
+\left\vert \phi _{dev}^{-}\right\rangle \left\vert \psi
_{cat}^{-}\right\rangle \right)  \label{sch2}
\end{equation}%
where%
\begin{equation}
\left\vert \psi _{cat}^{\pm }\right\rangle =\frac{1}{\sqrt{2}}\left(
\left\vert alive\right\rangle \pm \left\vert dead\right\rangle \right) ,
\label{cat}
\end{equation}%
we can see that if the measurement result of the device is $\left\vert \phi
_{dev}^{+}\right\rangle $, then the state of the cat becomes $\left\vert
\psi _{cat}^{+}\right\rangle $. Now if we have the projective measurement
operator%
\begin{equation}
P_{cat}^{+}=\left\vert \psi _{cat}^{+}\right\rangle \left\langle \psi
_{cat}^{+}\right\vert  \label{projector}
\end{equation}%
and apply it on the cat, the projection will be successful with probability $%
100\%$. This tells it apart from the classical mixture described by $\rho
_{cat}$ in Eq. (\ref{mixture}) because applying $P_{cat}^{+}$\ on $\rho
_{cat}$ will result in either $\left\vert \psi _{cat}^{+}\right\rangle $ or $%
\left\vert \psi _{cat}^{-}\right\rangle $\ with the equal probabilities $%
50\% $, since Eq. (\ref{mixture}) can also be expressed as%
\begin{equation}
\rho _{cat}=\frac{1}{2}\left\vert \psi _{cat}^{+}\right\rangle \left\langle
\psi _{cat}^{+}\right\vert +\frac{1}{2}\left\vert \psi
_{cat}^{-}\right\rangle \left\langle \psi _{cat}^{-}\right\vert .
\end{equation}%
Similarly, according to Eq. (\ref{sch2}), if the measurement result of the
device turned out to be $\left\vert \phi _{dev}^{-}\right\rangle $ instead,
then the state of the cat became $\left\vert \psi _{cat}^{-}\right\rangle $.
Applying $P_{cat}^{+}$ on it will always result in the failure of the
projection, which makes it differ from the classical mixture $\rho _{cat}$
too.

All in all, no matter the Schr\"{o}dinger's cat is presented as Eq. (\ref%
{sch}) or Eq. (\ref{cat}), with the proper projective operators $P_{Sch}^{+}$
or $P_{cat}^{+}$ we can discriminate it from the mixture Eq. (\ref{classical}%
) or Eq. (\ref{mixture}). Now the question is: do such operators actually
exist for real cats? We shall show below that the answer is negative, so
that it is impossible to directly observe and verify the existence of Schr%
\"{o}dinger's cats.

\section{Impossibility of observing real Schr\"{o}dinger's cats}

Applying our no-observing theorem on the Schr\"{o}dinger's cat paradox, we
can see that if the physical carrier is a real cat, then it is impossible to
find it as the Schr\"{o}dinger's cat state, no matter it is expressed as Eq.
(\ref{sch}) or Eq. (\ref{cat}). This is because, first, while there could be
medical cases standing on the blurred line between life and death which are
hard to define technically, they are obviously not what the Schr\"{o}%
dinger's cat paradox is talking about. Instead, most of the time
discriminating the state of a living cat from a dead one is not a hard task.
Secondly, though the secrets of life still have not been unfold completely,
we all know that bringing a dead cat back to life is impossible, at least
with current science. Therefore, the states $\left\vert alive\right\rangle $
and $\left\vert dead\right\rangle $\ satisfy the two conditions in the
theorem even if the laboratory is taken as the entire universe, so we can
take $\left\vert L\right\rangle =\left\vert alive\right\rangle $ and $%
\left\vert D\right\rangle =\left\vert dead\right\rangle $. Consequently, the
theorem tells us that the projective operator $P_{S}$ in Eq. (\ref{Ps}) does
not exist in reality. As a special case (by taking $a=b=1/\sqrt{2}$), it is
impossible to construct the projective operator $P_{cat}^{+}$ in Eq. (\ref%
{projector}). Otherwise, as shown in the proof of the theorem, we can start
with a real dead cat, then apply $P_{cat}^{+}$\ and $P_{L}$\ successively
and bring it back to life.

For the same reason, we can also take $\left\vert L\right\rangle =\left\vert
undecayed\right\rangle \left\vert alive\right\rangle $ and $\left\vert
D\right\rangle =\left\vert decayed\right\rangle \left\vert dead\right\rangle
$ instead, because no matter what the relationship between $\left\vert
undecayed\right\rangle $ and $\left\vert decayed\right\rangle $\ is, the
irreversibility from $\left\vert alive\right\rangle $ to $\left\vert
dead\right\rangle $\ already guarantees that the states $\left\vert
L\right\rangle $ and $\left\vert D\right\rangle $ defined this way meet the
requirement in the theorem. Then it immediately follows that the projective
operator $P_{Sch}^{+}$ in Eq. (\ref{project both}) does not exist either.

Not only the operators $P_{Sch}^{+}$ and $P_{cat}^{+}$ are absent, the
theorem also excludes the existence of other $P_{S}$ in Eq. (\ref{Ps}) for
any nonvanishing $a$ and $b$. Consequently, there will be no non-commuting bases. The only bases remain available physically are these corresponding to either $a=0$ or $%
b=0 $, e.g., $\{\left\vert undecayed\right\rangle \left\vert
alive\right\rangle ,\left\vert decayed\right\rangle \left\vert
dead\right\rangle \}$\ and $\{\left\vert alive\right\rangle ,\left\vert
dead\right\rangle \}$. But in these bases, the
Schr\"{o}dinger's cat described by Eq. (\ref{sch}) or Eq. (\ref{cat}) will
show no difference from the trivial classical mixtures Eq. (\ref{classical})
or Eq. (\ref{mixture}). As a result, even if there is something claimed to
be a real Schr\"{o}dinger's cat, no evidence can ever be found to support
this claim.

Note that Aaronson, Atia, and Susskind \cite{qi2269} also reached a similar
conclusion with a different approach. They proved that if there exists a
unitary operation which can discriminate $\left( \left\vert
alive\right\rangle +\left\vert dead\right\rangle \right) /\sqrt{2}$ from $%
\left( \left\vert alive\right\rangle -\left\vert dead\right\rangle \right) /%
\sqrt{2}$, then there will be a unitary operation which can turn $\left\vert
dead\right\rangle $ into $\left\vert alive\right\rangle $. Comparing with
their approach, our proof showed that if there exists a projective operator
which can discriminate $\left( \left\vert alive\right\rangle +\left\vert
dead\right\rangle \right) /\sqrt{2}$ from $\left( \left\vert
alive\right\rangle -\left\vert dead\right\rangle \right) /\sqrt{2}$, then $%
\left\vert dead\right\rangle $ can be turned into $\left\vert
alive\right\rangle $ via a sequence of projective measurements. Since
projective operations are not unitary, it seems that the two proofs
complement each other nicely.

\section{Relationship with quantum superpositions of lifeless systems}

In contrast to the above result, we all know that many quantum superpositions of
lifeless systems were proven possible to observe. To see why they have such
a divergence from real cats, let us use polarized photons as an example. Let
$\left\vert 0\right\rangle $\ ($\left\vert 1\right\rangle $) denote the
state of a photon with $0%
{{}^\circ}%
$\ ($90%
{{}^\circ}%
$) polarization, and define $\left\vert x^{\pm }\right\rangle =(\left\vert
0\right\rangle \pm \left\vert 1\right\rangle )/\sqrt{2}$. Again, suppose
that we have many copies of each states. If we put a polarizer with its
polarizing axis along the $0%
{{}^\circ}%
$\ direction and detect whether a photon can pass through it, then we are
performing the measurement in the basis $\{\left\vert 0\right\rangle
,\left\vert 1\right\rangle \}$. In this measurement, many states can pass
the polarizer with probability $50\%$, such as the pure states $\left\vert
x^{+}\right\rangle $ and $\left\vert x^{-}\right\rangle $, as well as the
equal mixture of $\left\vert 0\right\rangle $\ and $\left\vert
1\right\rangle $\ described by the density matrix%
\begin{equation}
\rho _{ph}=\frac{1}{2}\left\vert 0\right\rangle \left\langle 0\right\vert +%
\frac{1}{2}\left\vert 1\right\rangle \left\langle 1\right\vert .
\end{equation}

Now if there is a state claimed to be $\left\vert x^{+}\right\rangle $, how
can we tell it apart from $\rho _{ph}$? This can never be done if $%
\{\left\vert 0\right\rangle ,\left\vert 1\right\rangle \}$ is the only basis
in existence without other non-commuting bases. But for photons, $\left\vert x^{+}\right\rangle $\ actually
stands for $45%
{{}^\circ}%
$ polarized. We can rotate the axis of the polarizer to the $45%
{{}^\circ}%
$\ direction and detect whether the photon can pass through it. This is
performing the measurement in the basis $\{\left\vert x^{+}\right\rangle
,\left\vert x^{-}\right\rangle \}$, with the projective operators $%
\{\left\vert x^{+}\right\rangle \left\langle x^{+}\right\vert ,\left\vert
x^{-}\right\rangle \left\langle x^{-}\right\vert \}$. In this case, the
state $\left\vert x^{+}\right\rangle $\ will pass the polarizer with
certainty, while the mixture $\rho _{ph}$ will fail with probability $50\%$.
Thus the difference is shown.

This example tells us that the existence of $\left\vert x^{+}\right\rangle $%
\ can be verified once we can perform the measurement in the basis $%
\{\left\vert x^{+}\right\rangle ,\left\vert x^{-}\right\rangle \}$. But for
real cats, such a measurement is impossible because the theorem proved that
the corresponding projective operator $P_{cat}^{+}$ in Eq. (\ref{projector})
does not exist. Therefore, while it is not a secret that $45%
{{}^\circ}%
$ polarized photons indeed exist, the existence of real Schr\"{o}dinger's
cats can never be justified physically. The key difference, as seen in the
theorem, is that the evolution from $\left\vert alive\right\rangle $ to $%
\left\vert dead\right\rangle $ is irreversible for real cats. On the
contrary, for photons the two terms $\left\vert 0\right\rangle $\ and $%
\left\vert 1\right\rangle $\ in $\left\vert x^{+}\right\rangle =(\left\vert
0\right\rangle +\left\vert 1\right\rangle )/\sqrt{2}$\ can be converted to
each other with a rotation operation, which is a reversible unitary
transformation.

The same is true for the experimental implementations of the so-called
\textquotedblleft cat states\textquotedblright\ (not on real cats) \cite%
{qi2257,qi2258,qi2259,qi2261,qi2235,qi2248,qi2254,qi2256,qi2243,qi2244,qi2242}
and the quantum superpositions of certain \textquotedblleft macroscopically
distinguishable\textquotedblright\ states \cite{qi2228,qi2229}, as well as
various theoretical schemes for such experiments \cite%
{qi2221,qi2220,qi2251,qi2245,qi2219,qi2250,qi2249,qi2252,qi2265}, where the
polarized photons in the above example are replaced by either trapped ions,
cavity QED systems, superconducting qubits, or other physical carriers. In
these implementations, no matter the states are presented as $N(\left\vert
\alpha \right\rangle \left\vert \alpha \right\rangle +\left\vert -\alpha
\right\rangle \left\vert -\alpha \right\rangle )$ or $N(\left\vert \alpha
\right\rangle +\left\vert -\alpha \right\rangle )$, the two terms in the
superpositions are interconvertible to each other via unitary
transformations, which is always reversible. Therefore, their existence does
not conflict with our no-go theorem.

\section{Discussion}

Our above general proof shows that as long as irreversible processes are
involved, then a certain class of projective measurements cannot exist.
Consequently, unless we can raise the dead, or there can never be any
physically implementable process which can tell a real Schr\"{o}dinger's cat
apart from a classical mixture of ordinary cats.

Technically speaking, if the existence of something is not verifiable
physically, then we can safely say that this object does not exist. Just
like asking whether there is anything existing outside the Hubble radius of
the Universe, it does not have physical meanings because no matter the
answer is yes or no, it cannot make any physically observable difference on
our world, as long as the current basic laws of science still hold. In this
sense, the inexistence of the aforementioned measurements may also be
understood as the inexistence of real Schr\"{o}dinger's cats. More
generally, our theorem can be understood as: a necessary condition for two
states to form a quantum superposition is that the evolution between them
has to be reversible. Here the reversible process does not limited to
unitary transformations. Any process can be covered, even probabilistic ones.

Our result may also help us better understand the long-standing question why
the macroscopic world looks classical though the microscopic world is
quantum. It is not the size that matters. Instead, it depends on whether the
system is complicated enough to allow irreversible processes so that non-commuting measurement bases no longer exist.

\section*{Acknowledgements}

This work was supported in part by Guangdong Basic
and Applied Basic Research Foundation (Grant No. 2019A1515011048).



\end{document}